\documentclass[conference]{IEEEtran}
\IEEEoverridecommandlockouts
\usepackage{cite}
\usepackage{amsmath,amssymb,amsfonts}
\usepackage{algorithm}
\usepackage{algpseudocode}
\usepackage{graphicx}
\usepackage{float}
\usepackage{subfigure}
\usepackage{textcomp}
\usepackage{xcolor}
\usepackage{multirow}

\def\BibTeX{{\rm B\kern-.05em{\sc i\kern-.025em b}\kern-.08em
    T\kern-.1667em\lower.7ex\hbox{E}\kern-.125emX}}
\begin{document}

\bstctlcite{IEEEexample:BSTcontrol}
\title{Dynamic Gradient Sparse Update for Edge Training\\
}

\author{\IEEEauthorblockN{I-Hsuan Li and Tian-Sheuan Chang}
\IEEEauthorblockA{\textit{Institute of Electronics, National Yang Ming Chiao Tung University,}
Hsinchu, Taiwan
}
\IEEEauthorblockA{Email: \{shelly.ee11, tschang\}@nycu.edu.tw}
}

\maketitle

\begin{abstract}%

Training on edge devices enables personalized model fine-tuning to enhance real-world performance and maintain data privacy. However, the gradient computation for backpropagation in the training requires significant memory buffers to store intermediate features and compute losses. This is unacceptable for memory-constrained edge devices such as microcontrollers. To tackle this issue, we propose a training acceleration method using dynamic gradient sparse updates. This method updates the important channels and layers only and skips gradient computation for the less important channels and layers to reduce memory usage for each update iteration. In addition, the channel selection is dynamic for different iterations to traverse most of the parameters in the update layers along the time dimension for better performance. The experimental result shows that the proposed method enables an ImageNet pre-trained MobileNetV2 trained on CIFAR-10 to achieve an accuracy of 85.77\% while updating only 2\% of convolution weights within 256KB on-chip memory. This results in a remarkable 98\% reduction in feature memory usage compared to dense model training.

\end{abstract}

\section{Introduction}  
In recent years, there has been an increasing demand to train deep neural networks (DNNs) on edge devices due to the requirement for privacy protection, user-specific tasks, and real-world applications. However, unlike inference, on-device training needs to fine-tune a pre-trained model to adapt to new data. This is extremely challenging since edge devices, such as microcontrollers, have a limited on-chip SRAM buffer size like 256KB. \cite{lin2022device} has pointed out that the required memory footprint could be tens to hundreds MB even for a lightweight model like MobileNetV2. This large footprint is required to store the intermediate features and compute the gradient for backpropagation. Therefore, the reduction of the memory footprint is one of the most critical design challenges for resource-limited edge devices. Otherwise, these data have to be stored on external DRAM and will result in significant memory bandwidth and power consumption~\cite{lee2021overview}. 

There are several ways to train DNNs efficiently with less memory and power cost. Most studies \cite{lin2022device, kwon2023tinytrain} investigate layer or channel selection algorithms to find important parameters to update and skip the calculation of unimportant gradients due to the requirement for memory constraints. 
However, these approaches use target datasets to find important parameters, which is not practical. In a real scenario, the model shall adapt to the unknown incoming data distribution (though within the same application domain), which will not be known beforehand. Similar situations also occur for their model pruning. In addition, the parameters to be updated are fixed during the training, which limits the model performance. 

Addressing the above issues, this paper proposes a dynamic gradient sparse update strategy to fine-tune a model under a memory constraint like 256KB. This method can reduce the footprint of memory on the chip while achieving high performance. Our method first prunes the model under the dataset of the pre-trained model instead of the target dataset to fit real situations, which can be done offline on the cloud side. Then, on the edge device, our method uses a simple criterion that selects later stages first with a constant selection ratio to decide layers or channels to be updated based on observations of the training process. All of these are in the memory budget. Channel selection will be dynamically changed to traverse all parameters if possible during performance optimization training. The experimental results show that this dynamic sparse update method achieves 85.77\% accuracy for MobileNetV2 on CIFAR-10 by updating 2\% of convolution weights and saving 98\% of feature maps to hold compared to the original dense model.

\section{Related Work} 
\subsection{Model Compression}
Model compression has been proposed to reduce the amount of data and computation by generating data as zeros, allowing the hardware to save memory and speed up. Two important categories of model compression techniques are structured pruning and unstructured pruning. Structured pruning can be divided into filter pruning and channel pruning \cite{he2017channel}. This method can produce regular weight matrices and make the network more efficient and compact. However, structured pruning suffers from notable accuracy loss. For unstructured pruning \cite{gale2019state}, it converts small weights to zero, which makes the weight distribution of zeros random and finer-grained. Since the weight distribution is irregular, it makes skip-zero calculation more complex and power inefficient.

\subsection{Memory-Efficient Training}
Several existing works have attempted to reduce the cost of transfer learning and save memory usage as much as possible. Simple ways are to update the classifier layer \cite{sharif2014cnn}, update the convolution bias \cite{cai2020tinytl}, and update the normalization layer \cite{frankle2020training}. However, these approaches will encounter an accuracy loss that cannot be ignored, and their performance will have a certain upper limit due to limited learning capacity. Several methods have also attempted to optimize both memory and performance. For example, SparseUpdate \cite{lin2022device} proposes automated selection with the contribution analysis method to search for the update scheme and skip the computation of the less important layers and channels. TinyTrain \cite{kwon2023tinytrain} introduces a task-adaptive sparse update method to dynamically select the layer and channel to update based on a multi-objective criterion. However, these approaches use target datasets to find important parameters and only update these parameters to achieve the goal of reducing memory usage, which is unrealistic. It is impossible for us to know the dataset used before performing on-device training. Moreover, the methods to find updated parameters are time-consuming and complicated. 

\section{Methodology} 

\begin{figure}[t]
\centering
\includegraphics[width=1.0\linewidth]{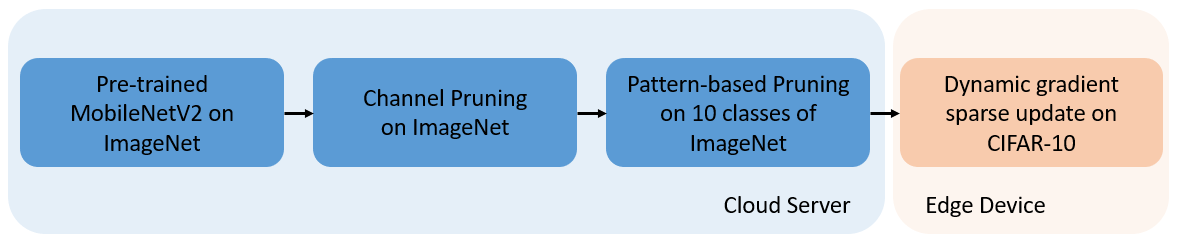}
\caption{The overall flow-chart with the network pruning and dynamic gradient sparse update procedure}
\label{network_procedure}
\end{figure}

We utilize pruning techniques to obtain a sparse model. 
However, the memory footprint is still too large for training on edge devices. To further reduce memory usage, we use our proposed dynamic gradient sparse update strategy to find the update scheme under memory constraints. Fig.~\ref{network_procedure} shows the flow of our proposed method. We first prune the model with the dataset of the pre-trained model. 
Then we apply the dynamic gradient sparse update strategy. For illustration purposes, we fine-tune the ImageNet pre-trained MobileNetV2 on the target CIFAR-10 dataset in the following description. The target memory constraint is 256KB.

\subsection{Pruning}

\subsubsection{Weight Pruning}
Channel pruning removes unnecessary output channels to accelerate convolution neural networks. To avoid significant performance degradation, we employ an efficient channel pruning method \cite{fang2023depgraph} that introduces the automatic method, Dependency Graph (DepGraph), to find the dependency between layers and the coupled parameters of the group for pruning. This way can ensure that all the removed parameters are consistently unimportant.
After employing channel pruning, we use the pattern-based pruning method \cite{niu2020patdnn} to further improve sparsity. This kind of semi-structured pruning offers higher compression rates compared to structured pruning techniques, such as channel pruning, while maintaining or even improving performance. 
We take the average of the absolute values of the weights for each convolution layer and then sort them. The higher layers are more sensitive to loss and should be pruned less. Therefore, we used the average values to determine the sparsity of each layer, and the sparsity of the filters of the same layer will be the same. This approach is beneficial in solving the processing element (PE) utilization problem since when different PEs compute the same layer but for different channels, the required computation time is expected to be similar, given the equal sparsity.

\subsubsection{Activation Pruning}
ReLU will produce the inherent sparsity of activation. However, the sparsity caused by ReLU is not high enough. To enhance sparsity and consider execution efficiency, we employ the block activation pruning method \cite{shih2020zebra}. If all the activation values in the same block are smaller than the threshold, all the activations will be pruned to zero. Since the accuracy will drop more with larger block size, we choose to use the block size as 2 in our experiments.

\subsection{Proposed Dynamic Gradient Sparse Update Strategy} 
Compared with inference accelerators, training accelerators need to hold more data, such as intermediate features and losses. However, there is no need for a training accelerator to store and compute values for the entire backpropagation process when doing transfer learning. Inspired by \cite{lin2022device}, we can skip the calculation of the gradients of the less important layers and channels to discard the corresponding output features for memory saving. In detail, we can update a subset of important weight channels in a layer, and the update ratio of the channels can be self-defined. This method prunes gradients instead of weights to save memory usage and computation.

\begin{figure}[t]
\centering
\includegraphics[width=0.8\linewidth]{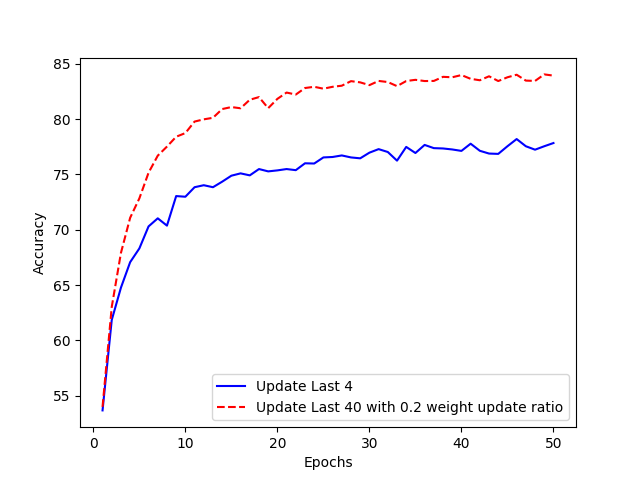}
\caption{Training process of updating last 4 layers and last 40 layers with 0.2 weight update ratios}
\label{last4_vs_last40}
\end{figure}

\subsubsection{Layer/Channel Selection}
To select important parameters, we observe that updating as many layers as possible can achieve higher performance. Fig.~\ref{last4_vs_last40} shows the training processes under 256KB extra memory usage in the backpropagation phase to update the last 4 layers and the last 40 layers with 0.2 weight update ratios. We can easily find that, under the same memory constraints, updating more layers gives much higher accuracy.
Moreover, it is better to update the later layers than to update the front layers. Because the front layers are for general features, updating will highly possible to destroy the model. Therefore, we choose small weight update ratios in later layers for channel updates to update as many layers as possible under memory constraints and freeze the front layers. This approach is very simple to determine the parameters to update for the image classification model and solves the complex layer/channel selection problem in \cite{lin2022device}.

\begin{algorithm}
\caption{Dynamic Gradient Sparse Update Strategy}\label{alg:alg1}
\begin{algorithmic}
\Require {Backbone weights $W$, Early fixed iterations $j$, Dynamic iterations $k$, Late fixed iterations $l$, Memory budgets $M$, Weight update ratio $r$}
\Ensure
\State Perform Layer/Channel Selection using $\{W,M,r\}$
\For{$t=1,...,j$}
\Comment{Fixed Stage}
\State Update the selected layers/channels of $W$
\EndFor
\For{$t=1,...,k$}
\Comment{Dynamic Stage}
\State Perform random channel selection for updated layer using $r$
\State Update the selected layers/channels of $W$
\EndFor
\For{$t=1,...,l$}
\Comment{Fixed Stage}
\State Update the selected layers/channels of $W$
\EndFor
\end{algorithmic}
\end{algorithm}

\subsubsection{Dynamic Strategy}
We find that if we dynamically change the update channels with random selection, the performance can increase more than if we use a fixed method. The proposed dynamic gradient sparse update strategy is shown in Algorithm~\ref{alg:alg1}. The rationale behind this design comes from the fact that during the initial stages of training, the model is still adapting to the training dataset. Consequently, the effect of frequently altering the channels to be updated is comparable to a method in which the channels are updated in a fixed manner. As training progresses over a certain period, a dynamic update strategy becomes more advantageous. During this dynamic training phase, it provides the model with an opportunity to update all its weights over time, granting it a higher likelihood of updating a greater number of weights compared to a fixed parameter updating approach. However, as the training process advances into its later stages and convergence is achieved, only minor fine-tuning is necessary. Consequently, the dynamic updating strategy exhibits diminishing returns, which makes it less effective. Therefore, in the later stages of training, the choice is made to opt for a fixed parameter updating strategy.

In summary, to save the overhead of the dynamic method, we design our training strategy that fixes the update scheme in the early stage, employs the dynamic method in the middle stage, and then fixes the update scheme again in the later stage. Contrary to existing training strategies that fix the update parameters. Our method can dynamically change the update channels to achieve the effect of traversing most of the parameters in the update layers in the time dimension.

\section{Experimental Results} 

\begin{table}[bp]
\centering
\caption{Detailed class mapping from ImageNet classes to CIFAR-10 classes \cite{huang2021unlearnable}}
\label{imagenet_10_class}
\begin{tabular}{c|c}
CIFAR-10 Class & ImageNet CLASS   \\ \hline
Airplane       & Airliner         \\
Automobile     & Wagon            \\
Bird           & Humming Bird     \\
Cat            & Siamese Cat      \\
Deer           & Ox               \\
Dog            & Golden Retriever \\
Frog           & Tailed Frog      \\
Horse          & Zebra            \\
Ship           & Container Ship   \\
Truck          & Trailer Truck   
\end{tabular}
\end{table}

\subsection{Settings}
To demonstrate the effectiveness of our method, we perform transfer learning tasks using the PyTorch framework. The following are the setups used in the experiments. We load the pre-trained MobileNetV2 on ImageNet and perform channel pruning \cite{fang2023depgraph}. Then, the pruning model is fine-tuned on the selected 10 classes of ImageNet, which are the classes mapping from ImageNet to CIFAR-10. Table~\ref{imagenet_10_class} shows the detailed classes \cite{huang2021unlearnable}. At the same time, we perform pattern-based pruning \cite{niu2020patdnn} on the model to get a higher sparsity result. 
To conduct transfer learning on CIFAR-10 with dynamic gradient sparse update strategy and block activation pruning \cite{shih2020zebra} using threshold 0.15, we re-scaled the images to 224 × 224 to be consistent with ImageNet. The model is then trained for 50 epochs using a single batch size, SGD with a momentum of 0, and an initial learning rate of 0.1 with a linear warm-up in the first five epochs and then the cosine annealing decay strategy.

For normalization layers, batch normalization (BN) is the most widely used normalization in vision tasks in the CNN architecture. However, for memory-constrained edge devices, BN is not suitable, since it requires a large batch size to calculate the estimation of training statistics \cite{masters2018revisiting}. To solve the problems, we replace BN with group normalization (GN) \cite{wu2018group}, which can apply smaller training batches because the running statistics during training are calculated independently for different inputs. 

\subsection{Network Pruning Result} 
We first use channel pruning \cite{fang2023depgraph} to prune MobileNetV2 parameters from 3.5M to 2.01M, which produces 57.39\% sparsity, and FLOPs from 0.32G reduce to 0.15G, a reduction of 46.44\% ( 2.15x decrease ). And then, combined with pattern-based pruning \cite{niu2020patdnn}, the sparsity can reach 92\% and even higher with some accuracy loss.

\subsection{Dynamic Gradient Sparse Update Experiment}

\begin{figure}[t]
\centering
\includegraphics[width=0.8\linewidth]{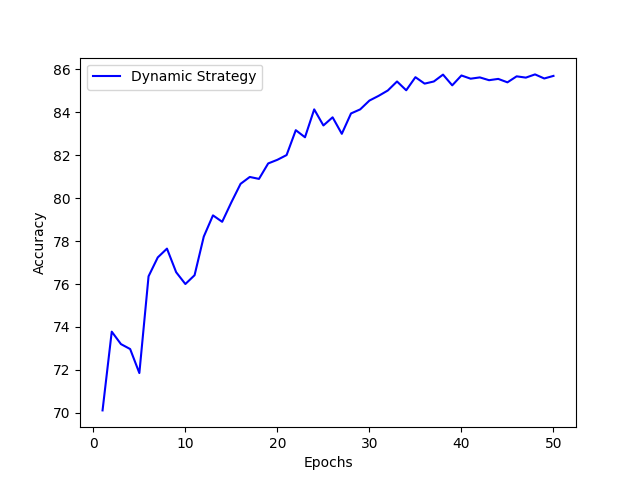}
\caption{Training process of dynamic strategy with 1 batch size}
\label{dynamic_acc}
\end{figure}

\begin{figure}[t]
\centering
\includegraphics[width=0.7\linewidth]{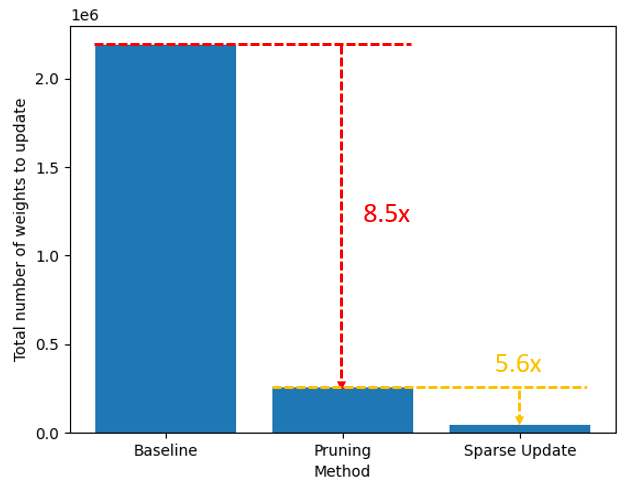}
\caption{Comparison of weights to update}
\label{parameter_update}
\end{figure}

Training of normalization layers involves complex operations, such as root and division, which causes significant overhead. Therefore, we choose to freeze the GN layers and reuse the mean and variance of the sparse pre-trained model in transfer learning. 

We update the last 42 layers with 0.2 weight update ratios on MobileNetV2 with sparsity 92\% on CIFAR-10. For epochs 1 to 10, we fix the update parameter. After 10 epochs, we randomly select 0.2 ratios of channels in the last 42 layers to be updated. Then, after 30 epochs, we fix the update parameter again as the final step. During the training process, additional memory usage includes activations and weights in backward propagation, all under 256KB. Fig.~\ref{dynamic_acc} shows the training process of the dynamic strategy experiment. Fig.~\ref{parameter_update} shows a comparison of the total number of weights to be updated in one epoch during different methods. Compared to the dense model baseline, our method only needs to update 2\% of convolution weights. 

\begin{table}[t]
\caption{Evaluation Results}
\label{accuracy}
\centering
\begin{tabular}{c|c|c}
Method         & Extra Memory & Accuracy \\ \hline
No Fine-tuning & 0MB          & 36.83\%  \\
Last           & Almost 0MB   & 59.34\%  \\
Full           & 2.5MB        & 90.33\%  \\
Fixed          & 0.25MB       & 84.3\%   \\
Dynamic (Ours) & 0.25MB       & 85.77\% 
\end{tabular}
\end{table}

\subsection{Evaluation}

Table~\ref{accuracy} presents the accuracy results by applying the pruned MobileNetV2 to CIFAR-10 with a fixed GN. This table compares our method to four cases: no fine-tuning, last only (update the final classifier layer), full (fine-tune the whole model), and fixed (fix the channel selection by our method). The accuracy will drop to 36.83\% if there is no fine-tuning. The full model fine-tuning has the highest accuracy, but also requires the largest memory footprint, 2.5MB, which exceeds the available 256KB on-chip memory. With our simple channel selection, even fixed selection can achieve an accuracy of 84. 3\%. This accuracy can be further improved by 1.47\% with our dynamic selection strategy. Both these two are within the 256KB constraint. 

Following the pruning process, we observe a noticeable loss of accuracy in the model. To improve performance, we use the knowledge distillation method \cite{hao2023vanillakd} to retrain accuracy after pruning. To facilitate a fair comparison with \cite{lin2022device}, we conduct experiments by training MobileNetV2 using a 100KB sparse update scheme. Our proposed method achieves an accuracy of 88.96\% on CIFAR-10. In contrast, \cite{lin2022device} produces an accuracy of 86.99\%. Our method outperforms \cite{lin2022device} by 1.97\%.

\subsection{Ablation Study and Analysis}
\subsubsection{Update Strategy}
In the determination of the update strategy, we explored an approach in which initially a larger number of layers were subject to updates, with each layer undergoing incremental weight updates. The decision on which layers to update was guided by evaluating the total sum of gradients for each layer. Layers with smaller total sums of gradients, indicating their relatively minor impact on loss, were considered candidates for freezing. The memory resources released from these frozen layers were reallocated to layers with larger total sums of gradients. This reallocation of resources allowed for an increased weight update ratio, facilitating a larger number of parameters to be updated within them.

However, it is worth noting that this approach did not consistently outperform a fixed-layer updating strategy. Variability in the relative total sums of gradients for each layer in different iterations was a challenge. This fluctuation resulted in situations where layers that were initially expected to receive more updates did not consistently have the opportunity to do so. As a result, the strategy did not consistently yield better results than a fixed-layer update approach.

\subsubsection{Normalization Layer}
We have tried to replace BN with multiple methods. One of the attempts is to replace normalization layers with a learnable scalarmultiplier $\alpha$, bias $\beta$, and a zero-centered noise injector $\delta$ \cite{liu2022nomorelization}. However, this method is not suitable for MobileNetV2 because it cannot be trained due to the Depthwise Convolution. Another attempt is to replace BN with FRN \cite{singh2020filter}. This approach can eliminate batch dependence in the training procedure and let us use a single batch size to train. However, the accuracy loss cannot be ignored.

\section{Conclusion}
In this paper, we propose a DNN training acceleration to reduce the memory footprint on the hardware. Our method first uses pruning methods to obtain a higher sparsity model. Then, we propose a dynamic gradient sparse update strategy to reduce the memory usage of output features and weights for computing gradients when backpropagation by skipping the computation of less important parameters.  Compared to the original dense model, we achieve 85.77\% accuracy for MobileNetV2 on CIFAR-10 by updating 2\% convolution weights and saving 98\% feature memory usage. In the future, we would like to design a memory-efficient training accelerator for transfer learning using our proposed method.

\bibliographystyle{IEEEtran}

\bibliography{IEEEabrv,bib/ieeeBSTcontrol,bib/thesis}

\end{document}